\documentclass[english,reprint,twocolumn,prl,amsmath,amssymb]{revtex4-1}
\usepackage[T1]{fontenc}
\usepackage[latin9]{inputenc}
\usepackage{amsthm}
\usepackage{graphicx}
\usepackage{braket}
\usepackage{bbm}
\usepackage{bm}
\usepackage[unicode=true, breaklinks=false, pdfborder={0 0 1}, backref=false, colorlinks=true, linkcolor=blue, urlcolor=blue, citecolor=blue]{hyperref}

\usepackage{standalone}

\makeatletter
 \theoremstyle{definition}
  \newtheorem{example}{\protect\examplename}

\allowdisplaybreaks

\newcommand{\tr}{\operatorname{tr}}

\let\originalleft\left
\let\originalright\right
\renewcommand{\left}{\mathopen{}\mathclose\bgroup\originalleft}
\renewcommand{\right}{\aftergroup\egroup\originalright}

\renewcommand\bra[1]{{\langle{#1}|}}
\makeatletter
\renewcommand\ket[1]{
  \@ifnextchar\bra{\k@t{#1}\!}{\k@t{#1}}
}
\newcommand\k@t[1]{{|{#1}\rangle}}
\makeatother
\makeatother

\usepackage{babel}
  \providecommand{\examplename}{Example}

\begin{document}

\title{Strong Quantum Darwinism and Strong Independence is equivalent to Spectrum Broadcast Structure}

\author{Thao P. Le}
\email{thao.le.16@ucl.ac.uk}

\affiliation{Dept. of Physics and Astronomy, University College London, Gower Street, London WC1E 6BT}

\author{Alexandra Olaya-Castro}
\email{a.olaya@ucl.ac.uk}

\affiliation{Dept. of Physics and Astronomy, University College London, Gower Street, London WC1E 6BT}
\begin{abstract}
How the objective everyday world emerges from the underlying quantum behaviour of its microscopic constituents is an open question at the heart of the foundations of quantum mechanics. Quantum Darwinism and spectrum broadcast structure are two different frameworks providing key insight into this question. Recent works, however, indicate these two frameworks can lead to conflicting predictions on the objectivity of the state of a system interacting with an environment.  Here we provide a resolution to this issue by defining strong quantum Darwinism and proving that it is equivalent to spectrum broadcast structure when combined with strong independence of the subenvironments. We further show that strong quantum Darwinism is sufficient and necessary to signal state objectivity without the requirement of strong independence. Our work unveils the deep connection between strong quantum Darwinism and spectrum broadcast structure, thereby making fundamental progress towards understanding and solving the emergence of classicality from the quantum world. Together they provide us a sharper understanding of the transition in terms of state structure, geometry, and quantum and classical information.
\end{abstract}
\maketitle
\begin{figure}
\begin{centering}
\includegraphics[width=0.75\columnwidth]{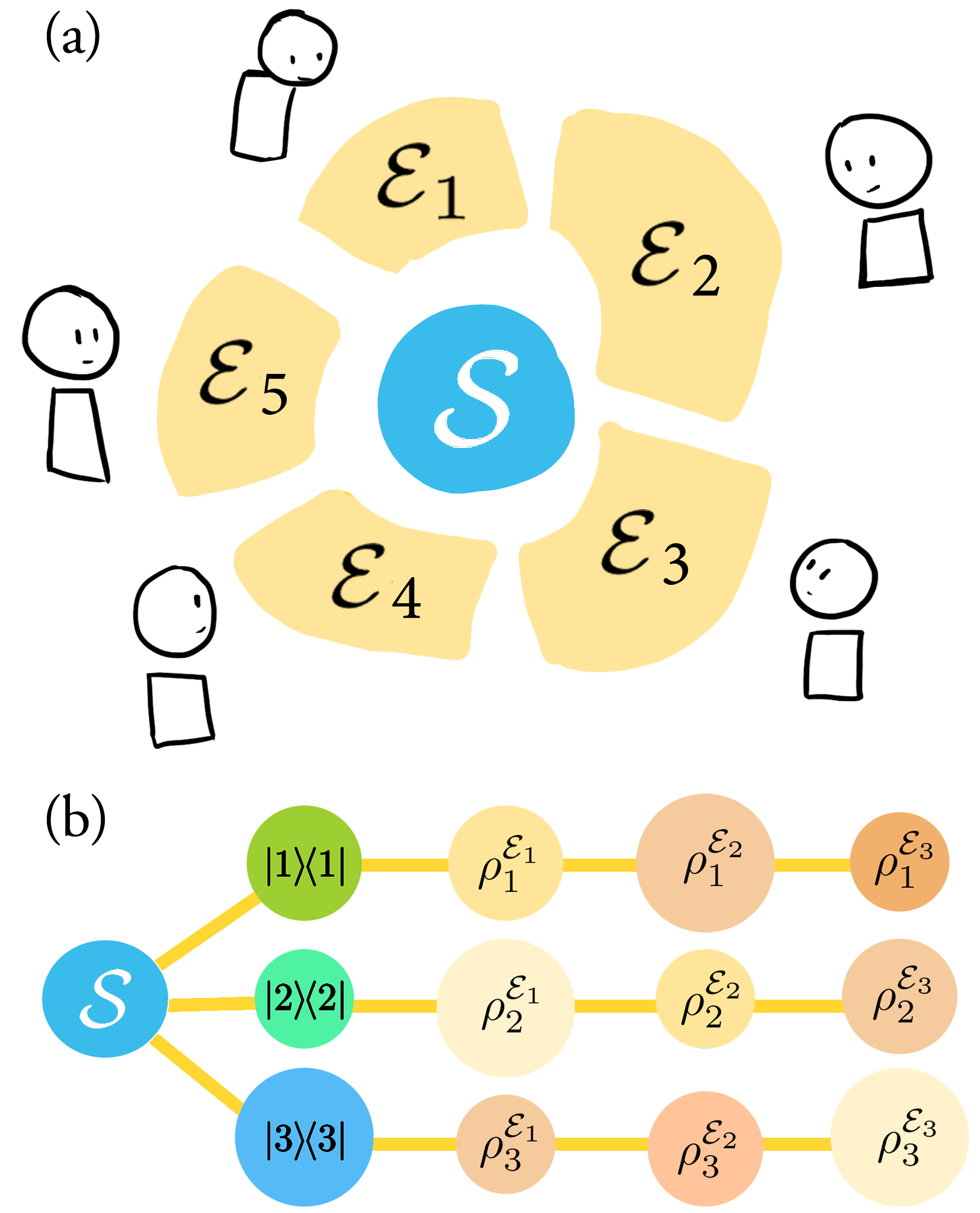}
\par\end{centering}
\caption{\textbf{Quantum Darwinism and spectrum broadcast structure}. \textbf{(a)} Quantum Darwinism recognises that the environment is made up of different fragments, for example $\mathcal{E}=\mathcal{E}_{1}\otimes\mathcal{E}_{2}\otimes\cdots\otimes\mathcal{E}_{5}$. Different observers access the properties of the system by measuring different environment fragments. \textbf{(b)} A spectrum broadcast structure state can be viewed as the existence of different branches, where there are different, distinguishable states $\left\{ \rho_{i}^{\mathcal{E}_k}\right\} _{i}$ given different conditional system states.\label{fig:Objectivity}}
\end{figure}

Through the interaction with large environments, quantum systems lose their underlying subjective quantum behaviour and appear objective to independent observers. How this transition from the quantum to the classical world happens is not fully understood. Decoherence theory takes a huge leap towards solving this problem \cite{Joos1985,Schlosshauer2005,Schlosshauer2007}, however, by itself does not explain other important aspects of objectivity such as the redundancy of information. Two different frameworks that can explain this information redundancy are quantum Darwinism \cite{Zurek2009} and spectrum broadcasting \cite{Horodecki2015}. They are illustrated in Fig. \ref{fig:Objectivity}. Quantum Darwinism divides the environment into multiple independent fragments and shows that the system-environment interaction can lead to information about the system state being duplicated into such fragments. Quantum Darwinism emerges when multiple different fragments have sufficient information about the system, measured using the quantum mutual information $I\left(\mathcal{S}:\mathcal{F}\right)$ between system $\mathcal{S}$ and fragment $\mathcal{F}$. Spectrum broadcasting uses a specific classical-quantum state structure called spectrum broadcast structure to signal the emergence of objectivity. These frameworks are complementary in their approach to signal objectivity. The former is entropic in nature, while the latter, focused on the state structure, is geometric. Both frameworks have been studied in various spin-spin and spin-boson models \cite{Pleasance2017, Balaneskovic2015,Balaneskovic2016,Ollivier2004,Blume-Kohout2005,Riedel2012,Zwolak2016,Zwolak2009,Zwolak2010,Giorgi2015,Zwolak2014,Zwolak2017,Lampo2017,Mironowicz2017a}, illuminated spheres \cite{Riedel2010,Riedel2011,Korbicz2014,Horodecki2015}, quantum Brownian motion \cite{Tuziemski2015,Tuziemski2015a,Tuziemski2016,Galve2016,Blume-Kohout2008,Paz2009}, single $N$-level environments \cite{Perez2010,Le2018}, generalised probabilistic theories \cite{Scandolo2018}, and even in QED \cite{Tuziemski2017}, gravitational \cite{Korbicz2017a}, and experimental quantum-dot \cite{Brunner2008, Brunner2010, Brunner2012, Burke2010, Ferry2015} and photonic \cite{Ciampini2018} settings. Together, quantum Darwinism and spectrum broadcast structure have made important conceptual contributions to the long-standing problem of the quantum-to-classical transition.

Quantum Darwinism and spectrum broadcast structure rely on an agreed definition of objectivity:

\emph{Definition: Objectivity.\textemdash }A system state is \emph{objective} if it is (1) simultaneously accessible to many observers (2) who can all determine the state independently \emph{without perturbing it} and (3) all arrive at the same result \cite{Ollivier2004,Zurek2009,Horodecki2015}.

For example, our observations of the moon are objective\textemdash by independently observing the light emitted by the moon, different observers can describe the same moon. The definition of what it means to be objective in-and-of-itself is up for debate (this definition can be thought of as \emph{inter-subjectivity} rather than objectivity \emph{per se} \cite{Mironowicz2017}), but that debate is not purpose of this Letter. For our purpose of understanding when and why quantum Darwinism can be inconsistent with emergence of classical objectivity, this basic definition is sufficient. 

Recent works have shown examples in which quantum Darwinism can falsely herald objectivity of the state of a quantum system~\cite{Horodecki2015,Pleasance2017,Le2018}. \citet{Pleasance2017} have considered a qubit coupled to a bosonic environment and found that the mutual information ``plateau'' that is traditionally used to signal quantum Darwinism\textemdash and thus objectivity\textemdash was in fact largely comprised of \emph{quantum discord} rather than classical information. We have investigated objectivity in a qubit interacting with an $N$-level environment and have shown that there can be a non-negligible amount of quantum discord in a situation where quantum Darwinism had apparently emerged \cite{Le2018}. \citet{Horodecki2015} have argued that certain entangled states could satisfy quantum Darwinism while not being objective. The existence of quantum discord in these cases means that the condition of ``measurement without perturbation'' will fail and hence the system state is \emph{not} objective, despite what quantum Darwinism suggests.

It is precisely in this respect that traditional quantum Darwinism and the spectrum broadcast structure diverge. Spectrum broadcast structure explicitly fulfills the requirement of non-perturbation of measurement in the sense of Bohr non-disturbance \cite{Bohr1935}. Furthermore, spectrum broadcast structure \emph{implies} quantum Darwinism, whereas the converse direction does not hold. The mutual information plateau condition of quantum Darwinism is not sufficient to determine whether a state is objective.

Here we propose a resolution to this issue by formulating a stronger version of quantum Darwinism\textemdash \emph{strong quantum Darwinism}. The original mutual information condition is replaced by a stronger condition using the accessible information, Holevo quantity, and the quantum discord. Rather than requiring sufficient mutual information, sufficient classical information, as given by the accessible information and Holevo quantity, is required. The quantum discord must also be vanishing\textemdash if the quantum discord is nonzero, then there is information about the system that is not locally accessible by the observer measuring their fragment \cite{Ollivier2001,Henderson2001,Modi2014}. We prove that strong quantum Darwinism is equivalent to spectrum broadcast structure when combined with strong independence. This leads to the corollary that strong quantum Darwinism is sufficient and necessary for objectivity. In contrast with spectrum broadcast structure, system objectivity does not require strong independence. In its mathematical simplicity, strong Quantum Darwinism makes fundamental progress towards understanding and solving the emergence of classicality from the quantum world. We also suggest an entropic measure for strong quantum Darwinism that complements the geometric distance bound for spectrum broadcast structure~\cite{Mironowicz2017}, unifying the various perspectives used to study the quantum-to-classical transition---state structure, geometric distances, and quantum information theory.

\emph{Quantum Darwinism.\textemdash }For a system-environment state $\ket{\Psi_{\mathcal{SE}}}$, the reduced density matrix of system is 
\begin{align}
\rho_{\mathcal{S}} & =\tr_{\mathcal{E}}\left[\ket{\Psi_{\mathcal{SE}}}\bra{\Psi_{\mathcal{SE}}}\right].
\end{align}
Decoherence theory and environment superselection \cite{Zurek1993,Zurek2003} describe the preferred pointer basis in which a quantum system decoheres. This is given by the pointer observable $\Pi_{\mathcal{S}}$, that the system will diagonalise under the influence of the environment \cite{Zwolak2013,Zurek1993,Zurek2000, Zurek2003}. Since the complete evolution is unitary, quantum Darwinism argues that information about the system $\mathcal{S}$\textemdash and in particular, information about the observable $\Pi_{\mathcal{S}}$\textemdash must be encoded somewhere in the environment $\mathcal{E}$. This holds for all quantum dynamics regardless of the details of the system-environment interaction \cite{Brandao2015, Knott2018}. Observers then obtain information about the system by measuring fragments of the environment. The von Neumann entropy of the system $H\left(\mathcal{S}\right)=H\left(\rho_{\mathcal{S}}\right)=-\tr\rho_{\mathcal{S}}\log_{2}\rho_{\mathcal{S}}=-\sum_i p_i \log_{2}p_i$, where $p_i$ are its eigenvalues, characterises the information contained within the system. The system is \emph{objective} from the perspective of the observer if they can obtain precisely this information from their measurement of their environment fragment without perturbing the system. Here, we take the mathematical definition of Bohr-nondisturbance \citep{Bohr1935}: the state remains unchanged after discarding the measurement results of the POVM $\{ M_{i}^{\mathcal{E}_{k}}\}$ on any sub-environment $\mathcal{E}_{k}$: $\sum_{i} M_{i}^{\mathcal{E}_{k}}\rho_{\mathcal{SE}} {M_{i}^{\mathcal{E}_{k}}}^{\dagger}=\rho_{\mathcal{SE}}$.

Suppose an observer has access to some fragment $\mathcal{F}$ of the environment. The reduced density matrix of system and fragment is
\begin{align}
\rho_{\mathcal{SF}} & =\tr_{\mathcal{E}\backslash\mathcal{F}}\left[\ket{\Psi_{\mathcal{SE}}}\bra{\Psi_{\mathcal{SE}}}\right],
\end{align}
where the trace is over all the environment $\mathcal{E}$ except the fragment $\mathcal{F}$.
The quantum mutual information, $I\left(\mathcal{S}:\mathcal{F}\right)=H\left(\mathcal{S}\right)+H\left(\mathcal{F}\right)-H\left(\mathcal{SF}\right)$, measures the total quantum and classical correlations between $\mathcal{S}$ and $\mathcal{F}$.
The classically accessible information is $I_{acc}\left(\mathcal{S}:\mathcal{F}\right)$, and for an objective state, this is equal to the Holevo information on the fragment  $\mathcal{F}$ conditioned on the system $\mathcal{S}^{\Pi}$ after measurement of  $\Pi_{\mathcal{S}}$ on the system \cite{Zwolak2013},
\begin{align}
\chi\left(\mathcal{S}^{\Pi}:\mathcal{F}\right) &= \max_{\hat{\Pi}_{\mathcal{S}}}\left\{ H\left(\sum_{a}p_{a}\rho_{\mathcal{F}|a}\right)-\sum_{a}p_{a}H\left(\rho_{\mathcal{F}|a}\right)\right\},
\end{align}
where $a$ are the measurement results of a system POVM $\hat{\Pi}_{\mathcal{S}}$, $p_{a}$ is the probability of that result and $\rho_{\mathcal{F}|a}$ is the conditional state on the fragment. The Holevo information bounds the maximum information about the classical random variable on the system with probabilities $\{p_a\}$ that can be determined via measurements of the fragment $\mathcal{F}$ \cite{Holevo1973}.
The quantum discord \cite{Ollivier2001,Henderson2001},
\begin{align}
\mathcal{D}\left(\mathcal{S}^{\Pi}:\mathcal{F}\right) &= H(\mathcal{S})-H(\mathcal{SF})+\min_{\hat{\Pi}_{\mathcal{S}}} H\left(\sum_{a}p_{a}\rho_{\mathcal{F}|a}\right),
\end{align}
describes quantum (\emph{i.e.} non-classical) correlations beyond entanglement: whilst entanglement cannot be prepared using local operations and classical communication (LOCC), quantum discord cannot be measured using LOCC \cite{Modi2014}.

From the complementarity between the classical information in the Holevo quantity with the quantum information given by the quantum discord, the quantum mutual information between system and fragment is in fact $I\left(\mathcal{S}:\mathcal{F}\right)=\chi\left(\mathcal{S}^{\Pi}:\mathcal{F}\right)+\mathcal{D}\left(\mathcal{S}^{\Pi}:\mathcal{F}\right)$ \cite{Zwolak2013}. We have the components required to state strong Quantum Darwinism:

\emph{Definition: Strong Quantum Darwinism.\textemdash }A system state $\mathcal{S}$ is said to be objective when there exists a fragment of the environment $\mathcal{F}\subseteq\mathcal{E}$ such that the following condition holds:
\begin{align}
I\left(\mathcal{S}:\mathcal{F}\right)=I_{acc}\left(\mathcal{S}:\mathcal{F}\right)=\chi\left(\mathcal{S}^{\Pi}:\mathcal{F}\right) & =H\left(\mathcal{S}\right),\label{eq:strong_QD}
\end{align}
where $I\left(\mathcal{S}:\mathcal{F}\right)$ is the quantum mutual information, $I_{acc}\left(\mathcal{S}:\mathcal{F}\right)$ is the accessible information, $\chi\left(\mathcal{S}^{\Pi}:\mathcal{F}\right)$ is the Holevo quantity in the pointer basis $\Pi$ and $H\left(\mathcal{S}\right)$ is the von Neumann entropy of the system. For the system state to be objective, Eq.~\eqref{eq:strong_QD} must also hold for multiple disjoint sub-fragments $\mathcal{F}_i$ corresponding to multiple independent observers, where $\mathcal{F}=  \mathcal{F}_{1} \cup\mathcal{F}_{2}\cup\ldots\cup\mathcal{F}_{k} $, $\mathcal{F}_{i}\cap\mathcal{F}_{j}=\emptyset$ for all $i\neq j$.

Strong quantum Darwinism recognises the difference between shared classical information and shared quantum information: shared quantum information may have nonzero discord and hence information that is not locally accessible. In contrast, traditional quantum Darwinism \cite{Zurek2009} only requires that $I\left(\mathcal{S}:\mathcal{F}\right)=H\left(\mathcal{S}\right)$, and it was \emph{assumed} that large majority of that information would be classical in nature. However, as the studies of Refs. \cite{Pleasance2017, Le2018,Horodecki2015} show, some of that information can be explicitly quantum in nature. The stronger condition $I\left(\mathcal{S}:\mathcal{F}\right)=\chi\left(\mathcal{S}^{\Pi}:\mathcal{F}\right)$ is also assumed by \citet{Zwolak2013}, which they call ``surplus decoherence'', but was not rigorously imposed as a core part of quantum Darwinism as we have done here.

To define the redundancy and spread of the classical information, suppose there are fragments $\mathcal{F}_{\delta}$ with size $\left|\mathcal{F}_{\delta}\right|=f_{\delta}\left|\mathcal{E}\right|$ that contain classical information
\begin{align}
I\left(\mathcal{S}:\mathcal{F}\right)&\approx\chi\left(\mathcal{S}^{\Pi}:\mathcal{F}\right)\geq\left(1-\delta\right)H\left(\mathcal{S}^{\Pi}\right)
\end{align}
that is approximately the information about the system. The redundancy $R_{\delta}$ is the number of unique copies of that information, \emph{i.e.}, the number of disjoint fragments $\mathcal{F}_{\delta,i}$ (where $i$ indexes different fragments) that contain that approximate information, $R_{\delta}=\left|\mathfrak{F}_{\delta}\right|$, where
\begin{align}
\mathfrak{F}_{\delta} & =\left\{ \mathcal{F}_{\delta,i}\left|\begin{array}{c}
\chi\left(\mathcal{S}^{\Pi}:\mathcal{F}_{\delta,i}\right)\geq\left(1-\delta\right)H\left(\mathcal{S}^{\Pi}\right),\\
\mathcal{F}_{\delta,i}\cap\mathcal{F}_{\delta,j}=\emptyset\,\forall\,i\neq j
\end{array}\right.\right\} .
\end{align}
This is bounded by the minimum fraction size $f_{\delta,min}$: $R_{\delta}\leq1/f_{\delta,min}$. If $\chi\left(\mathcal{S}^{\Pi}:\mathcal{F}\right)\geq\left(1-\delta\right)H\left(\mathcal{S}^{\Pi}\right)$, then the discord is bounded by $\mathcal{D}\left(\mathcal{S}^{\Pi}:\mathcal{F}\right)\leq\delta H\left(\mathcal{S}^{\Pi}\right)$.

\begin{figure}
\begin{centering}
\includegraphics[width=0.9\columnwidth]{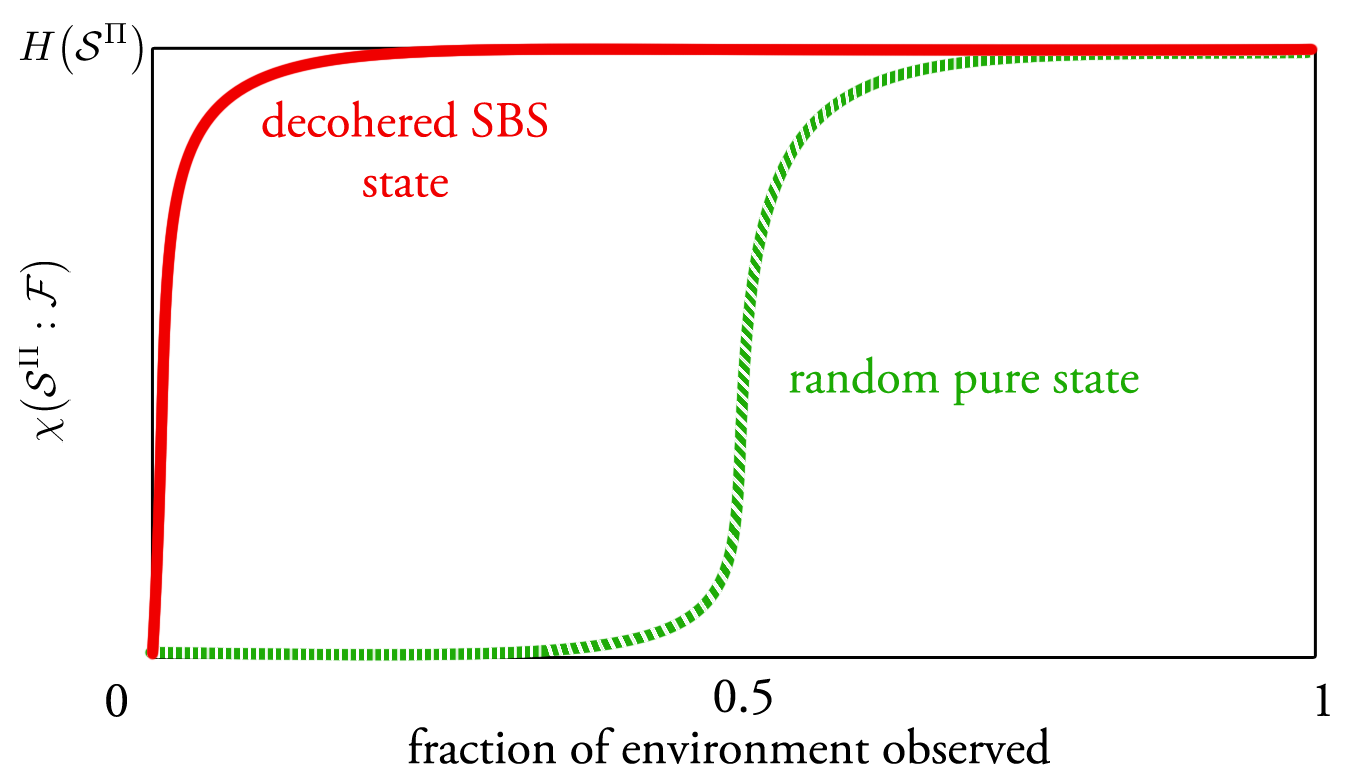}
\par\end{centering}
\caption{\textbf{Accessible classical information given by the Holevo information $\chi\left(\mathcal{S}^{\Pi}:\mathcal{F}\right)$ about the system stored in different fractions of the environment.} (Green dashed) For pure states picked out at random (through a Haar measure), the amount of accessible classical information and quantum discord are typically roughly the same. (Red solid) For a state that satisfies strong quantum Darwinism, such as the reduced GHZ state, $\chi\left(\mathcal{S}^{\Pi}:\mathcal{F}\right)$ will be approximately equal to the system entropy $H\left(\mathcal{S}\right)$ for small fractions of the environment, suggesting the existence of multiple copies of the information and hence redundancy. (c.f. analogous Figure 2 in \citet{Zurek2009}.) \label{fig:Refined-Quantum-Darwinism}}
\end{figure}

A rapid rise of the classical information $\chi\left(\mathcal{S}^{\Pi}:\mathcal{F}\right)$, as shown schematically in Fig. \ref{fig:Refined-Quantum-Darwinism}, implies that only a small fraction $f_{\delta}$ of the environment is required to have access to all the information in the system and hence suggesting that there is a large redundancy $R_{\delta}\approx1/f_{\delta}$; this occurs for post-decohered system-environment states (that have spectrum broadcast structure). In contrast, Haar-random pure system-environment states \footnote{For example, a Haar-random pure state {$| \psi_{\mathcal{SE}}\rangle = U | \psi_0\rangle$} can be realised using some reference state {$|\psi_0 \rangle$} and a random unitary matrix {$U$} drawn with a uniform probability from the set of all unitary matrices on the system and environment.} will tend to have a mixture of classical and quantum correlations between any system and fragment, and a fairly large fraction would be required to access any substantial amount of information about the system.

Now, we present the definitions for spectrum broadcast structure and strong independence \cite{Horodecki2015}:

\emph{Definition: Spectrum Broadcast Structure.\textemdash }The joint state $\rho_{\mathcal{SF}}$ of the system $\mathcal{S}$ and a collection of subenvironments $\mathcal{F}=\mathcal{E}_{1}\otimes\ldots\otimes\mathcal{E}_{F}$ has spectrum broadcast structure if it can be written as:
\begin{align}
\rho_{\mathcal{SF}} & =\sum_{i}p_{i}\ket{i}_{\mathcal{S}}\bra{i}\otimes\rho_{i}^{\mathcal{E}_{1}}\otimes\cdots\otimes\rho_{i}^{\mathcal{E}_{F}},\label{eq:SBS}
\end{align}
where $\left\{ \ket{i}\right\} $ is the pointer basis, $p_{i}$ are probabilities, and all states $\rho_{i}^{\mathcal{E}_{k}}$ are perfectly distinguishable: $\rho_{i}^{\mathcal{E}_{k}}\rho_{j}^{\mathcal{E}_{k}}=0\,\forall\,i\neq j,$ for each observed environment $\mathcal{E}_{k}\in\left\{ \mathcal{E}_{1},\ldots,\mathcal{E}_{F}\right\} $ \cite{Horodecki2015}.

\emph{Definition: Strong independence.\textemdash }Subenvironments $\mathcal{E}_j\in\left\{ \mathcal{E}_{1},\mathcal{E}_{2},\ldots,\mathcal{E}_{F}\right\}$ have strong independence relative to the system $\mathcal{S}$ if their conditional mutual information is vanishing:
\begin{align}
I\left(\mathcal{E}_j : \mathcal{E}_k | \mathcal{S} \right) &=0 \quad \forall j\neq k.
\end{align}

Strong independence means that there are no correlations between the environments \emph{conditioned} on the information about the system.

With strong Quantum Darwinism, strong independence and spectrum broadcast structure now defined, we can prove the titular theorem of this Letter:

\emph{Theorem.\textemdash }A state $\rho_{\mathcal{SF}}$ has spectrum broadcast structure if and only if it satisfies strong quantum Darwinism and has strong independence.

The complete proof is given in Appendix A in the Supplemental Material \cite{supplemental}. In the forward direction $\left(\implies\right)$, an explicit form of the state $\rho_{\mathcal{SF}}$ with spectrum broadcast structure is used to calculate the various mutual and accessible information as well as the entropies required to fulfill strong quantum Darwinism. Spectrum broadcast structure explicitly satisfies strong independence~\cite{Horodecki2015}. In the reverse direction $\left(\Longleftarrow\right)$, the conditions of strong quantum Darwinism (including surplus decoherence of Ref. \cite{Zwolak2013}, and classical-only correlations of Ref. \cite{Piani2008}) enforce particular structures on the state $\rho_{\mathcal{SF}}$, and we use these properties and general properties of states, to recover bipartite spectrum broadcast structure, $\rho_{\mathcal{SF}} =\sum_{i}p_{i}\ket{i}_{\mathcal{S}}\bra{i}\otimes\rho_{i}^{\mathcal{F}}$, where $\rho_{i}^{\mathcal{F}}$ are mutually distinguishable. Applying strong independence recovers the form in Eq.~\eqref{eq:SBS}.

\emph{Corollary 1.\textemdash }Strong Quantum Darwinism is equivalent to bipartite spectrum broadcast structure.

This proof comes from the proof of main theorem: the result has been encapsulated into a Corollary in order to prove the following statement about objectivity:

\emph{Corollary 2.\textemdash }Strong Quantum Darwinism is sufficient and necessary for objectivity:
\begin{equation}
\left(\begin{array}{c}
\text{strong}\\
\text{quantum Darwinism}
\end{array}\right)\iff\begin{array}{c}
\left(\text{objectivity}\right).
\end{array}\label{eq:SQD_objectivity}
\end{equation}

\emph{Proof.\textemdash }\citet{Horodecki2015} prove precisely Eq. (\ref{eq:SQD_objectivity}) for bipartite spectrum broadcast structure (in place of strong quantum Darwinism).
By Corollary 1, Eq. (\ref{eq:SQD_objectivity}) holds for strong quantum Darwinism.\hfill$\blacksquare$

\emph{Remark.\textemdash }In fact,  \citet{Horodecki2015} proved that objectivity $+$ strong independence $\implies$ spectrum broadcast structure. Hence, spectrum broadcast structure is an overly stringent requirement for objectivity, in contrast to strong quantum Darwinism which is both sufficient and necessary without additives. Strong independence itself is not required for the system objectivity: in Appendix B of the Supplemental Material \cite{supplemental}, we provide examples of objective states that exhibit bipartite broadcast structure, but not strong independence.

\emph{Measuring strong Quantum Darwinism.\textemdash }Through the lens of strong Quantum Darwinism, a large Holevo quantity $\chi\left(\mathcal{S}^{\Pi}:\mathcal{F}\right)$ is required for objectivity; whilst the discord $\mathcal{D}\left(\mathcal{S}^{\Pi}:\mathcal{F}\right)$ is a hindrance. As such, we suggest the following as a measure that captures the degree of objectivity of a state in the presence of discord:
\begin{align}
M^{\text{SQD}}\left(\rho_{\mathcal{SF}}\right) & \equiv \dfrac{H\left(\mathcal{S} \right) -\chi\left(\mathcal{S}^{\Pi}:\mathcal{F}\right)+\mathcal{D}\left(\mathcal{S}^{\Pi}:\mathcal{F}\right)}{2 H\left(\mathcal{S}\right)},
\end{align}
which takes values between $[0,1]$. Objectivity occurs when the minimum value is obtained, $M^{\text{SQD}}\left(\rho_{\mathcal{SF}}\right)=0$ signaling perfect strong quantum Darwinism. This measure is not unique, implying a certain partial ordering of states. The components of Eq. \eqref{eq:strong_QD} can be combined to form other valid measures with different orderings.

\citet{Mironowicz2017} have defined a geometric distance bound on how close a state $\rho_{\mathcal{SF}}$ is to being of spectrum broadcast structure:
\begin{align}
T^{\text{SBS}}\left(\rho_{\mathcal{SF}}\right) & =\dfrac{1}{2}\min_{\Pi_{\mathcal{S}}}\left\Vert \rho_{\mathcal{SF}}-\rho_{\mathcal{S}^{\Pi}\mathcal{F}}^{\text{SBS}}\right\Vert _{1}\leq\eta\left(\rho_{\mathcal{SF}}\right),
\end{align}
where
\begin{equation}
\begin{split}\eta\left(\rho_{\mathcal{SF}}\right) & \equiv\left\Vert \rho_{\mathcal{SF}}-\rho_{\mathcal{S}^{\Pi}\mathcal{F}}\right\Vert _{1}\\
 & \qquad+\sum_{i\neq j}\sqrt{p_{i}p_{j}}B\left(\rho_{\mathcal{F}|i},\rho_{\mathcal{F}|j}\right).
\end{split}
\end{equation}
The fidelity is $B\bigl(\rho_{1},\rho_{2}\bigr)=\left\Vert \sqrt{\rho_{1}}\sqrt{\rho_{2}}\right\Vert _{1}$, and $\rho_{\mathcal{S}^{\Pi}\mathcal{F}}=\sum_{i}p_{i}\ket{i}\bra{i}\otimes\rho_{\mathcal{F}|i}$ is the post-measurement (separable) state. Notice that our proposed $M^{\text{SQD}}\left(\rho_{\mathcal{SF}}\right)$ and \citet{Mironowicz2017}'s $\eta\left(\rho_{\mathcal{SF}}\right)$ are related to each other. The term $\left\Vert \rho_{\mathcal{SF}}-\rho_{\mathcal{S}^{\Pi}\mathcal{F}}\right\Vert _{1}=\mathcal{D}_{geo}\left(\mathcal{S}^{\Pi}:\mathcal{F}\right)$ is the \emph{geometric} quantum discord \cite{Dakic2010}, hence it is related to the entropic quantum discord. For two-qubit states, we have the explicit bound with the entropic quantum discord: $\left\Vert \rho_{\mathcal{SF}}-\rho_{\mathcal{S}^{\Pi}\mathcal{F}}\right\Vert _{1}\geq\sqrt{2}\mathcal{D}\left(\mathcal{S}^{\Pi}:\mathcal{F}\right)\geq\mathcal{D}\left(\mathcal{S}^{\Pi}:\mathcal{F}\right)$ \cite{Girolami2011,Girolami2011a,Paula2013,Roga2016}. Similarly, for a qubit system, the Holevo quantity is bounded as $\chi\left(\mathcal{S}^{\Pi}:\mathcal{F}\right)\geq I_{acc}\left(\mathcal{S}:\mathcal{F}\right)\geq H\left(\left\{ p_{1},p_{2}\right\} \right)-2\sqrt{p_{1}p_{2}}B\left(\rho_{\mathcal{F}|1},\rho_{\mathcal{F}:2}\right)$ (Eq. (5) in \cite{Jain2006}). Hence, in the case where the system-fragment $\rho_{\mathcal{SF}}$ is a two-qubit state,
\begin{align}
\eta\left(\rho_{\mathcal{SF}}\right) & \geq\mathcal{D}\left(\mathcal{S}^{\Pi}:\mathcal{F}\right)-\chi\left(\mathcal{S}^{\Pi}:\mathcal{F}\right)+H\left(\left\{ p_{1},p_{2}\right\} \right)\\
 & =2 H\left(\mathcal{S}\right)M^{\text{SQD}}\left(\rho_{\mathcal{SF}}\right).
\end{align}

The calculation of the entropic quantities of strong quantum Darwinism requires optimisation over measurements on the system. Without the use of the computable bound $\eta(\rho_\mathcal{SF})$, calculating the distance to the set of spectrum broadcasting states would require optimisation over both the system and all the subenvironments bases.

\emph{Discussion.\textemdash }We have shown a fundamental shift in understanding the emergence of classicality through \emph{classical} information redundancy as opposed to more general quantum information redundancy.
We introduced strong quantum Darwinism by identifying that shared \emph{classical} information is required for objectivity and by noting that the existence of quantum correlations hinders objectivity. Formally, we examined the nature of the shared system-environment information using the tools of quantum information theory. By proving that the combined strong quantum Darwinism and strong independence is equivalent to spectrum broadcast structure, we have provided a sharper understanding of the quantum-to-classical transition: strong quantum Darwinism alone is necessary and sufficient for objectivity of a system state, capturing succinctly the minimal requirements of objectivity. In contrast, spectrum broadcast structure describes both objectivity of the system state and partial objectivity of the environment states.
Finally, we have suggested a possible measure for the degree of objectivity using classical and quantum information, complementing the state structure and geometric perspectives of spectrum broadcasting.

The discrepancy between the classical accessible information and quantum mutual information observed in earlier papers \cite{Pleasance2017, Le2018} is now resolved: the discrepancy implies that strong quantum Darwinism does not emerge, and there is no objectivity nor spectrum broadcast structure. Strong quantum Darwinism also addresses the concerns by \citet{Horodecki2015}, whereby traditional quantum Darwinism emerges even when the system-environment state was clearly entangled. The example in \cite{Horodecki2015} is the following:
\begin{align}
\rho_{\mathcal{SE}} & =pP_{\left(a\ket{00}+b\ket{11}\right)}+\left(1-p\right)P_{\left(a\ket{10}+b\ket{01}\right)},
\end{align}
where $P_{\ket{\psi}}=\ket{\psi}\bra{\psi}$, $p\neq1/2$, $a=\sqrt{p}$ and $b=\sqrt{1-p}$. In Appendix C \cite{supplemental}, we determine that $I\left(\mathcal{S}:\mathcal{E}\right)=H\left(\mathcal{S}\right)$, \emph{i.e.}. quantum Darwinism is satisfied, whilst $\mathcal{X}\left(\mathcal{S}^{\Pi}:\mathcal{E}\right)\neq H\left(\mathcal{S}\right)$ for $p\neq 0,1$. Strong Quantum Darwinism is not satisfied in general, and this is consistent with the correct conclusion that the system is not objective.

Strong Quantum Darwinism opens up yet further questions to be addressed. We have shown that strong quantum Darwinism deviates from spectrum broadcast structure when there are intra-sub-environmental correlations. Then, when does strong quantum Darwinism deviate from the usual quantum Darwinism?
Many past studies of models in the literature find quantum Darwinism also assume no system self-Hamiltonian, or that the system Hamiltonian $H_S$ commutes with the coupling Hamiltonian $H_I$. In contrast, the two examples where strong quantum Darwinism is needed \cite{Pleasance2017,Le2018} both have that $[H_S,H_I]\neq 0$ yet with very different kinds of system-bath Hamiltonians. The work of Ref. \cite{Hossein-Nejad2015} shows that the commuting properties of $H_S$ and $H_I$ can shape the nature of the correlations, work, entropy \emph{etc.} in bipartite systems. These examples also displayed strong system-environment correlations and non-Markovian dynamics. Therefore, we conjecture that strong quantum Darwinism deviates from traditional quantum Darwinism when there is a non-negligible self Hamiltonian, a coupling Hamiltonian that does not commute with it, and strong system-environment correlations.

The quantum-to-classical transition remains an unsolved problem. Strong quantum Darwinism captures, formally and conceptually, the essence of what will be required:  the emergence of perfect classical correlations and the disintegration of quantum correlations between objective objects and independent observers.

Upon completion of our manuscript, two experimental works have been reported investigating quantum Darwinism in photonic \cite{Chen2018} and spin \cite{Unden2018} environments, which base their analysis in the Holevo information. Our work and the newly introduced concept of Strong Quantum Darwinism gives solid foundation to these experimental works.

We thank R. Horodecki, J. Korbicz, and P. Horodecki for discussions. We thank the anonymous referees for their constructive feedback. This work was supported by the Engineering and Physical Sciences Research Council {[}grant number EP/L015242/1{]}.

\bibliographystyle{apsrev4-1}
\bibliography{biblio}

\newpage
\onecolumngrid

\appendix

\section*{Supplemental Material for ``Strong Quantum Darwinism and Strong Independence is equivalent
to Spectrum Broadcast Structure''}

\section{Proof of the equivalence between spectrum broadcast structure and combined strong quantum Darwinism and strong independence (main Theorem)}

\subsection{Proof $\left(\protect\implies\right)$: spectrum broadcast structure implies strong quantum Darwinism and strong independence}

First, we prove that if $\rho_{\mathcal{SF}}$ has spectrum broadcast structure, then we have strong quantum Darwinism, with $I\left(\mathcal{S}:\mathcal{F}\right)=I_{acc}\left(\mathcal{S}:\mathcal{F}\right)=\chi\left(\mathcal{S}^{\Pi}:\mathcal{F}\right)=H\left(\mathcal{S}^{\Pi}\right)=H\left(\mathcal{S}\right)$.

The most general spectrum broadcast structure state has form
\begin{align}
\rho_{\mathcal{SF}}^{\text{SBS}} & =\sum_{i}p_{i}\ket{i}\bra{i}\otimes\rho_{i}^{\mathcal{E}_{1}}\otimes\cdots\otimes\rho_{i}^{\mathcal{E}_{F}},
\end{align}
where $\left\{ \ket{i}\right\} $ is the pointer basis, $p_{i}$ are probabilities, and the conditional states $\rho_{i}^{\mathcal{E}_{k}}$ are perfectly distinguishable: $\rho_{i}^{\mathcal{E}_{k}}\rho_{j}^{\mathcal{E}_{k}}=0$ for all $i\neq j$. Conditioned on system state $\ket{i}$, the conditional environment states are product and hence satisfy strong independence.

The condition of perfect distinguishability means that $\left\{ \rho_{i}^{\mathcal{E}_{k}}\right\} _{i}$ have mutually orthogonal supports and are simultaneously diagonalisable. Usually, the accessible information is bounded by the Holevo information: $I_{acc}\left(\mathcal{S}:\mathcal{F}\right)\leq\chi\left(\mathcal{S}^{\Pi}:\mathcal{F}\right)$, but since the terms $\left\{ \rho_{i}^{\mathcal{E}_{1}}\otimes\cdots\otimes\rho_{i}^{\mathcal{E}_{F}}\right\} _{i}$ commute, we have that $I_{acc}\left(\mathcal{S}:\mathcal{F}\right)=\mathcal{\chi}\left(\mathcal{S}^{\Pi}:\mathcal{F}\right)$ \cite{Ruskai2002}. 

Without loss of generality, we can define the eigenbasis for these fragment states as $\left\{ \ket{j}_{\mathcal{E}_{k}}\right\} _{j}$ for each subenvironment $\mathcal{E}_{k}$, such that the eigendecomposition of each state, conditioned on the system being in state $\ket{i}$, can be written as 
\begin{align}
\rho_{i}^{\mathcal{E}_{k}} & =\sum_{j}d_{\mathcal{E}_{k}}\left(j,i\right)\ket{j}_{\mathcal{E}_{k}}\bra{j},
\end{align}
for suitable coefficients $d_{\mathcal{E}_{k}}\left(j,i\right)$: $\sum_{j}d_{\mathcal{E}_{k}}\left(j,i\right)=1$ for all $k$ and all $i$, and $d_{\mathcal{E}_{k}}\left(j,i\right)$ take values such that $\rho_{i}^{\mathcal{E}_{k}}\rho_{j}^{\mathcal{E}_{k}}=0$ for all $i\neq j$. Hence,
\begin{align}
\rho_{\mathcal{F}|i} & =\sum_{j_{1}}d_{1}\left(j_{1},i\right)\ket{j_{1}}\bra{j_{1}}\otimes\cdots\otimes\sum_{j_{F}}d_{f}\left(j_{F},i\right)\ket{j_{F}}\bra{j_{F}}\\
 & =\sum_{j_{1},\ldots,j_{F}}d_{1}\left(j_{1},i\right)\cdots d_{F}\left(j_{F},i\right)\ket{j_{1}}\otimes\cdots\otimes\ket{j_{F}}\bra{j_{1}}\otimes\cdots\otimes\bra{j_{F}}\\
 & =\sum_{\boldsymbol{j}}d\left(\boldsymbol{j},i\right)\ket{\boldsymbol{j}}\bra{\boldsymbol{j}},
\end{align}
where $\boldsymbol{j}=\left(j_{1},\ldots,j_{F}\right)$ and we have defined $d\left(\boldsymbol{j},i\right)\equiv d_{1}\left(j_{1},i\right)\cdots d_{F}\left(j_{F},i\right)$. Each of the $\ket{\boldsymbol{j}}$ are orthonormal, and so we have an eigenbasis $\left\{ \ket{\boldsymbol{j}}\right\} _{\boldsymbol{j}}$. Therefore
\begin{align}
I_{acc}\left(\mathcal{S}:\mathcal{F}\right) & =\mathcal{\chi}\left(\mathcal{S}^{\Pi}:\mathcal{F}\right)=H\left(\sum_{i}p_{i}\rho_{\mathcal{F}|i}\right)-\sum_{i}p_{i}H\left(\rho_{\mathcal{F}|i}\right),\\
 & =H\left(\sum_{i}p_{i}\sum_{\boldsymbol{j}}d\left(\boldsymbol{j},i\right)\ket{\boldsymbol{j}}\bra{\boldsymbol{j}}\right)-\sum_{i}p_{i}H\left(\sum_{\boldsymbol{j}}d\left(\boldsymbol{j},i\right)\ket{\boldsymbol{j}}\bra{\boldsymbol{j}}\right)\\
 & =H\left(\sum_{\boldsymbol{j}}\left(\sum_{i}p_{i}d\left(\boldsymbol{j},i\right)\right)\ket{\boldsymbol{j}}\bra{\boldsymbol{j}}\right)-\sum_{i}p_{i}H\left(\left\{ d\left(\boldsymbol{j},i\right)\right\} _{\boldsymbol{j}}\right)\\
 & =H\left(\left\{ \sum_{i}p_{i}d\left(\boldsymbol{j},i\right)\right\} _{\boldsymbol{j}}\right)-\sum_{i}p_{i}H\left(\left\{ d\left(\boldsymbol{j},i\right)\right\} _{\boldsymbol{j}}\right)\\
 & =-\sum_{\boldsymbol{j}}\left(\sum_{i}p_{i}d\left(\boldsymbol{j},i\right)\right)\log_{2}\left(\sum_{k}p_{k}d\left(\boldsymbol{j},k\right)\right)-\sum_{i}p_{i}\left(-\sum_{\boldsymbol{j}}d\left(\boldsymbol{j},i\right)\log_{2}d\left(\boldsymbol{j},i\right)\right)\\
 & =-\sum_{i}p_{i}\sum_{\boldsymbol{j}}d\left(\boldsymbol{j},i\right)\left[\log_{2}\left(\sum_{k}p_{k}d\left(\boldsymbol{j},k\right)\right)-\log_{2}d\left(\boldsymbol{j},i\right)\right].
\end{align}

Due to the mutually orthogonal supports, $d_{\mathcal{E}_{e}}\left(j_{e},i\right)d_{\mathcal{E}_{e}}\left(j_{e},i^{\prime}\right)=0$ for $i\neq i^{\prime}$, \emph{i.e.,} for a given $j_{e}$, $d_{\mathcal{E}_{e}}\left(j_{e},i\right)$ is nonzero for \emph{at most one $i$}: $d_{\mathcal{E}_{e}}\left(j_{e},i\right)\neq0$ if and only if $j_{e}\in J_{i}^{\mathcal{E}_{e}}$, where $J_{i}^{\mathcal{E}_{e}}\cap J_{i^{\prime}}^{\mathcal{E}_{e}}=\emptyset$ are disjoint sets. 

Hence, $d\left(\boldsymbol{j},i\right)\neq0$ if and only if $\boldsymbol{j}\in J_{i}$, where $\left\{ J_{i}\right\} _{i}$ are disjoint sets $J_{i}\cap J_{i^{\prime}}=\emptyset$, where $J_i = J_i^{\mathcal{E}_1} \times J_i^{\mathcal{E}_2} \times \cdots \times J_i^{\mathcal{E}_F}$. Hence,
\begin{align}
I_{acc}\left(\mathcal{S}:\mathcal{F}\right) & =-\sum_{i}p_{i}\sum_{\boldsymbol{j}\in J_{i}}d\left(\boldsymbol{j},i\right)\left[\log_{2}\left(\sum_{k}p_{k}d\left(\boldsymbol{j},k\right)\right)-\log_{2}d\left(\boldsymbol{j},i\right)\right]\\
 & =-\sum_{i}p_{i}\sum_{\boldsymbol{j}\in J_{i}}d\left(\boldsymbol{j},i\right)\left[\log_{2}\left(p_{i}d\left(\boldsymbol{j},i\right)\right)-\log_{2}d\left(\boldsymbol{j},i\right)\right]\\
 & =-\sum_{i}p_{i}\sum_{\boldsymbol{j}\in J_{i}}d\left(\boldsymbol{j},i\right)\log_{2}p_{i}\\
 & =-\sum_{i}p_{i}\log_{2}p_{i}\\
 & =H\left(\left\{ p_{i}\right\} _{i}\right)=H\left(\mathcal{S}\right),
\end{align}
where $\sum_{\boldsymbol{j}\in J_{i}}d\left(\boldsymbol{j},i\right)=1$ by normalisation of the state.

Now, for $I\left(\mathcal{S}:\mathcal{F}\right)=H\left(\mathcal{S}\right)+H\left(\mathcal{F}\right)-H\left(\mathcal{SF}\right)$, note that
\begin{align}
H\left(\mathcal{SF}\right) & =H\left(\sum_{i}p_{i}\ket{i}\bra{i}\otimes\rho_{\mathcal{F}|i}\right)\\
 & =H\left(\sum_{i}p_{i}\ket{i}\bra{i}\sum_{\boldsymbol{j}}d\left(\boldsymbol{j},i\right)\ket{\boldsymbol{j}}\bra{\boldsymbol{j}}\right)\\
 & =H\left(\sum_{i}\sum_{\boldsymbol{j}}p_{i}d\left(\boldsymbol{j},i\right)\ket{i}\ket{\boldsymbol{j}}\bra{i}\bra{\boldsymbol{j}}\right)\\
 & =H\left(\left\{ p_{i}d\left(\boldsymbol{j},i\right)\right\} _{i,\boldsymbol{j}}\right)\\
 & =-\sum_{i}\sum_{\boldsymbol{j}}p_{i}d\left(\boldsymbol{j},i\right)\log_{2}\left(p_{i}d\left(\boldsymbol{j},i\right)\right)\\
 & =H\left(\mathcal{F}\right),
\end{align}
hence $I\left(\mathcal{S}:\mathcal{F}\right)=H\left(\mathcal{S}\right)$.

Finally, given the product form of $\rho_{\mathcal{F}|i}=\rho_{i}^{\mathcal{E}_{1}}\otimes\cdots\otimes\rho_{i}^{\mathcal{E}_{F}}$, if $F>1$, then the strong quantum Darwinism condition holds for multiple subenvironments $I\left(\mathcal{S}:\mathcal{E}_{k}\right)=I_{acc}\left(\mathcal{S}:\mathcal{E}_{k}\right)=\chi\left(\mathcal{S}^{\Pi}:\mathcal{E}_{k}\right)=H\left(\mathcal{S}\right)$, $k=1,\ldots,F$, and multiple different observers are able to measure their own fragment $\mathcal{E}_{k}$ without Bohr-disturbing \cite{Bohr1935} the system \cite{Horodecki2015}.\hfill$\blacksquare$

\subsection{Proof $\left(\protect\Longleftarrow\right)$ : strong quantum Darwinism and strong independence implies spectrum broadcast structure}

Now, we prove that if a state $\rho_{\mathcal{SF}}$ satisfies $I\left(\mathcal{S}:\mathcal{F}\right)=I_{acc}\left(\mathcal{S}:\mathcal{F}\right)=\chi\left(\mathcal{S}^{\Pi}:\mathcal{F}\right)=H\left(\mathcal{S}^{\Pi}\right)$, and if the state also has strong independence $I\left(\mathcal{E}_j : \mathcal{E}_k | \mathcal{S} \right) =0$ for all $j\neq k$, then it must have spectrum broadcast structure.

\subsubsection{Subproof: strong quantum Darwinism implies bipartite spectrum broadcast structure}

If $I_{acc}\left(\mathcal{S}:\mathcal{F}\right)=\chi\left(\mathcal{S}^{\Pi}:\mathcal{F}\right)$, then the ensemble $\left(p_{i},\rho_{\mathcal{F}|i}\right)$ of the state $\rho_{\mathcal{F}}$ must have commuting $\rho_{\mathcal{F}|i}$ elements \cite{Ruskai2002}. Therefore, they have a common eigenbasis $\ket{\psi_{j}}$, where $\rho_{\mathcal{F}|i}=\sum_{j}c\left(j,i\right)\ket{\psi_{j}}\bra{\psi_{j}}$ for some eigenvalues $\left\{ c\left(j,i\right)\right\} _{j}$. Then, $I_{acc}\left(\mathcal{S}:\mathcal{F}\right)=\chi\left(\mathcal{S}^{\Pi}:\mathcal{F}\right)=H\left(\mathcal{S}\right)$ implies that
\begin{align}
H\left(\left\{ p_{i}\right\} _{i}\right) & =H\left(\sum_{i}p_{i}\sum_{j}c\left(j,i\right)\ket{\psi_{j}}\bra{\psi_{j}}\right)-\sum_{i}p_{i}H\left(\sum_{j}c\left(j,i\right)\ket{\psi_{j}}\bra{\psi_{j}}\right)\\
-\sum_{i}p_{i}\log_{2}p_{i} & =-\sum_{j}\left(\sum_{i}p_{i}c\left(j,i\right)\right)\log_{2}\left(\sum_{k}p_{k}c\left(j,k\right)\right)-\sum_{i}p_{i}\left(-\sum_{j}c\left(j,i\right)\log_{2}c\left(j,i\right)\right) \label{eq:A24}\\
 & =-\sum_{i}p_{i}\sum_{j}c\left(j,i\right)\left\{ \log_{2}\left(\sum_{k}p_{k}c\left(j,k\right)\right)-\log_{2}c\left(j,i\right)\right\}. \label{eq:A25}
\end{align}

If \eqref{eq:A25} holds for every $i$, then it will hold for the average. Furthermore, the conditional states $\rho_{\mathcal{F}|i}$ always correctly predict the system state $\ket{i}$ for any probability distribution $\{p_i\}$, hence \eqref{eq:A25} should hold for every $i$:
\begin{align}
\log_{2}p_{i} & =\sum_{j}c\left(j,i\right)\left\{ \log_{2}\left(\sum_{k}p_{k}c\left(j,k\right)\right)-\log_{2}c\left(j,i\right)\right\} .
\end{align}
For this to hold, we require that for a given $j$, $c\left(j,k\right)$ is nonzero for a single $k$.
This is a correct solution by inspection. We can solve the above, as follows:
\begin{align}
\log_{2}p_{i} & =\sum_{j\text{ s.t. }c\left(j,i\right)\neq0}c\left(j,i\right)\log_{2}\left(\dfrac{\sum_{k}p_{k}c\left(j,k\right)}{c\left(j,i\right)}\right)\\
 & =\log_{2}\left[\prod_{j\text{ s.t. }c\left(j,i\right)\neq0}\left(\dfrac{\sum_{k}p_{k}c\left(j,k\right)}{c\left(j,i\right)}\right)^{c\left(j,i\right)}\right]\\
\implies p_{i} & =\prod_{j\text{ s.t. }c\left(j,i\right)\neq0}\left[\dfrac{\sum_{k}p_{k}c\left(j,k\right)}{c\left(j,i\right)}\right]^{c\left(j,i\right)}.
\end{align}
Since $\sum_{i}p_{i}=1$, we can write $p_{i}=1-\sum_{k\neq i}p_{k}$,
and for general probability distributions, we can consider $\left\{ p_{k}\right\} _{k\neq i}$
as a set of independent variables: these equations for the environment $\mathcal{F}$ should hold regardless of particular probabilities $\{p_i\}$. Hence,
\begin{align}
1-\sum_{k\neq i}p_{k} & =\prod_{j\text{ s.t. }c\left(j,i\right)\neq0}\left[\dfrac{p_{i}c\left(j,i\right)}{c\left(j,i\right)}+\dfrac{\sum_{k\neq i}p_{k}c\left(j,k\right)}{c\left(j,i\right)}\right]^{c\left(j,i\right)}\\
 & =\prod_{j\text{ s.t. }c\left(j,i\right)\neq0}\left[1-\sum_{k\neq i}p_{k}\left(1-\dfrac{c\left(j,k\right)}{c\left(j,i\right)}\right)\right]^{c\left(j,i\right)}.\label{eq:midstep}
\end{align}
First, if there was only one $j$ such that $c\left(j,i\right)\neq0$,
then $c\left(j,i\right)=1$ (since $\sum_{j}c\left(j,i\right)=1$)
and Eq. (\ref{eq:midstep}) reduces to 
\begin{align}
1-\sum_{k\neq i}p_{k} & =1-\sum_{k\neq i}p_{k}\left(1-c\left(j,k\right)\right)
\end{align}
and by equating the coefficients of $p_{k\neq i}$ on either side,
$1=1-c\left(j,k\right)$ hence, given $j$, $c\left(j,k\right)=0$ for all $k\neq i$.

Now, suppose there is two or more $j$'s such that $c\left(j,i\right)\neq0$, which we denote as
$\left\{ j_{1},\ldots,j_{n}\right\} =J_{i}$. Suppose $\sum_{k\neq i}p_{k}=p_{k_{1}}+p_{k_{2}}+\cdots+p_{k_{m}}$
(\emph{i.e.} the sum contains $m$ terms). Using the generalised multinomial
theorem, the term in the product can be written as:
\begin{align}
\left[1-\sum_{k\neq i}p_{k}\left(1-\dfrac{c\left(j,k\right)}{c\left(j,i\right)}\right)\right]^{c\left(j,i\right)}
 & \equiv\sum_{r_{1},\ldots,r_{m}=0}^{\infty}f\left(j,i,\vec{r}\right)p_{k_{1}}^{r_{1}}p_{k_{2}}^{r_{2}}\cdots p_{k_{m}}^{r_{m}}
\end{align}
where we have defined the following for simplicity:
\begin{equation}
\begin{split}f\left(j,i,\vec{r}\right)\equiv & \left(-1\right)^{r_{1}+\cdots+r_{m}}\dbinom{c\left(j,i\right)}{r_{1}+\cdots+r_{m}}\dbinom{r_{1}+\cdots+r_{m}}{r_{2}+\cdots+r_{m}}\cdots\dbinom{r_{m-1}+r_{m}}{r_{m}}\\
 & \times\left(1-\dfrac{c\left(j,k_{1}\right)}{c\left(j,i\right)}\right)^{r_{1}}\left(1-\dfrac{c\left(j,k_{2}\right)}{c\left(j,i\right)}\right)^{r_{2}}\cdots\left(1-\dfrac{c\left(j,k_{m}\right)}{c\left(j,i\right)}\right)^{r_{m}}.
\end{split}
\end{equation}
Now, Eq. (\ref{eq:midstep}) can be expanded to:
\begin{align}
1-\sum_{k\neq i}p_{k} & =\prod_{j\in J_{i}}\sum_{r_{1},\ldots,r_{m}=0}^{\infty}f\left(j,i,\vec{r}\right)p_{k_{1}}^{r_{1}}p_{k_{2}}^{r_{2}}\cdots p_{k_{m}}^{r_{m}}\\
 & =\sum_{r_{1}\left(j_{1}\right),\ldots,r_{m}\left(j_{1}\right)=0}^{\infty}f\left(j_{1},i,\vec{r}\left(j_{1}\right)\right)p_{k_{1}}^{r_{1}\left(j_{1}\right)}p_{k_{2}}^{r_{2}\left(j_{1}\right)}\cdots p_{k_{m}}^{r_{m}\left(j_{1}\right)}\times\sum_{r_{1}\left(j_{2}\right),\ldots,r_{m}\left(j_{2}\right)=0}^{\infty}f\left(j_{2},i,\vec{r}\left(j_{2}\right)\right)p_{k_{1}}^{r_{1}\left(j_{2}\right)}p_{k_{2}}^{r_{2}\left(j_{2}\right)}\cdots p_{k_{m}}^{r_{m}\left(j_{2}\right)}\nonumber \\
 & \phantom{=\sum_{r_{1}\left(j_{1}\right),\ldots,r_{m}\left(j_{1}\right)=0}^{\infty}}\times\cdots\times\sum_{r_{1}\left(j_{n}\right),\ldots,r_{m}\left(j_{n}\right)=0}^{\infty}f\left(j_{n},i,\vec{r}\left(j_{n}\right)\right)p_{k_{1}}^{r_{1}\left(j_{n}\right)}p_{k_{2}}^{r_{2}\left(j_{n}\right)}\cdots p_{k_{m}}^{r_{m}\left(j_{n}\right)}\\
 & =\sum_{\vec{r}\left(j_{1}\right),\vec{r}\left(j_{2}\right),\ldots,\vec{r}\left(j_{n}\right)=0}^{\infty}f\left(j_{1},i,\vec{r}\left(j_{1}\right)\right)f\left(j_{2},i,\vec{r}\left(j_{2}\right)\right)\cdots f\left(j_{n},i,\vec{r}\left(j_{n}\right)\right)\nonumber \\
 & \phantom{=\sum_{\vec{r}\left(j_{1}\right),\vec{r}\left(j_{2}\right),\ldots,\vec{r}\left(j_{n}\right)=0}^{\infty}}\times p_{k_{1}}^{r_{1}\left(j_{1}\right)+r_{1}\left(j_{2}\right)+\cdots+r_{1}\left(j_{n}\right)}p_{k_{2}}^{r_{2}\left(j_{1}\right)+r_{2}\left(j_{2}\right)+\cdots+r_{2}\left(j_{n}\right)}\cdots p_{k_{m}}^{r_{m}\left(j_{1}\right)+r_{m}\left(j_{2}\right)+\cdots+r_{m}\left(j_{n}\right)}.\label{eq:midstep2}
\end{align}

Now we equate the coefficients of the $p_{k}$ on either side of Eq.
(\ref{eq:midstep2}): 
\begin{align}
p_{k_{1}}:-1 & =f\left(j_{1},i,r_{1}=1\right)+f\left(j_{2},i,r_{1}=1\right)+\cdots+f\left(j_{2},i,r_{1}=1\right),\qquad(\text{all other }r=0)\\
 & =-c\left(j_{1},i\right)+c\left(j_{1},k_{1}\right)-c\left(j_{2},i\right)+c\left(j_{2},k_{1}\right)-\cdots-c\left(j_{n},i\right)+c\left(j_{n},k_{1}\right)\\
 & =-\sum_{j\in J_{i}}c\left(j,i\right)+\sum_{j\in J_{i}}c\left(j,k_{1}\right)\\
 & =-1+\sum_{j\in J_{i}}c\left(j,k_{1}\right)
\end{align}
since $\sum_{j\in J_{i}}c\left(j,i\right)=\sum_{j}c\left(j,i\right)=1$.
Hence, we must have $\sum_{j\in J_{i}}c\left(j,k_{1}\right)=0.$ This
is true for all the $k\neq i$, so
\begin{align}
\sum_{j\text{ s.t. }c\left(j,i\right)\neq0}c\left(j,k\right) & =0\qquad\forall k\neq i.\label{eq:final}
\end{align}
Since all the terms $0\leq c\left(j,i\right)\leq1$, the only possible solution to Eq. (\ref{eq:final}) is that all $c\left(j,k\right)=0$
for all $k\neq i$ and $j$ such that $c\left(j,i\right)\neq0$. That is, $c(j,k)$ is nonzero for only a single $k$.

Alternatively, the steps \eqref{eq:A24} to \eqref{eq:final} can be simplified by noting that \eqref{eq:A24} is exactly equal to $H(I) = H(J)-H(J|I)$ for the classical probability distribution $p_{I,J}(i,j) = p_i c(j,i)$. Since we can write $H(J|I) = H(I|J) - H(I)+H(J)$, \eqref{eq:A24} then becomes $H(I|J)=0$, meaning that knowing $J=j$ fixes the value of $I=i$.

This means that the state $\ket{\psi_{k}}_{\mathcal{F}}$ only appears when we condition on a specific, single system state $\ket{i}$, which implies that different conditional states are distinguishable:
\begin{align}
\rho_{\mathcal{F}|i}\rho_{\mathcal{F}|k} & =\sum_{j}c\left(j,i\right)c\left(j,k\right)\ket{\psi_{j}}\bra{\psi_{j}}\\
 & =\sum_{j}c\left(j,i\right)c\left(j,i\right)\delta_{ik}\ket{\psi_{j}}\bra{\psi_{j}}\\
 & =0,
\end{align}
for all $i\neq k$. And hence we have spectrum broadcast structure for the post-measurement state $\rho_{\mathcal{S}^{\Pi}\mathcal{F}}=\sum_{i}p_{i}\ket{i}\bra{i}\otimes\rho_{\mathcal{F}|i}$.

Recall that strong quantum Darwinism also has the condition that $I\left(\mathcal{S}:\mathcal{F}\right)=\chi\left(\mathcal{S}^{\Pi}:\mathcal{F}\right)$. This condition is called ``surplus decoherence'' by \citet{Zwolak2013}\textemdash note that our contribution is to make this a core part of strong quantum Darwinism, rather than assuming that it occurs. It implies that the original state has the classical-quantum branching form $\sum_{i}p_{i}\ket{i}\bra{i}\otimes\rho_{\mathcal{F}|i}$, \emph{i.e.}, $\rho_{\mathcal{SF}}=\rho_{\mathcal{S}^{\Pi}\mathcal{F}}$ has bipartite spectrum broadcasting structure. 

Note that \citet{Piani2008} have also proven the correspondence of $I\left(\mathcal{S}:\mathcal{F}\right)=I_{acc}\left(\mathcal{S}^{\Pi}:\mathcal{F}\right)$ (where the quantum mutual information is equal to classical mutual information) with classical-classical $\mathcal{SF}$ states $\rho_\mathcal{SF}=\sum_{i,j} p_{ij} \ket{i}\bra{i}\otimes\ket{\psi_j}\bra{\psi_j}$.

\subsubsection{Subproof: Recovering (multipartite) spectrum broadcast structure with strong independence}

Technically, we do not need to decompose $\mathcal{F}$ into smaller components in order to satisfy spectrum broadcast structure, as spectrum broadcast structure makes no explicit requirement on the number of observed sub-environments. We could, for example, consider $\mathcal{F}$ as a macrofraction \cite{Mironowicz2017}.

If one wished to decompose $\mathcal{F}$, then we need to enforce full strong Quantum Darwinism and strong independence. From full strong quantum Darwinism, multiple observers must be able to measure disjoint fractions of the fragment $\mathcal{F}$. Hence, there exist subfragments $\mathcal{F}=\mathcal{E}_{1}\otimes\cdots\otimes\mathcal{E}_{F}$ ($F>1$) for which the strong quantum Darwinism condition holds, and therefore $\rho_{\mathcal{S}\mathcal{E}_{k}}=\sum_{i}p_{i}\ket{i}\bra{i}\otimes\rho_{\mathcal{E}_{k}|i}$, where each $\left\{ \rho_{\mathcal{E}_{k}|i}\right\}_{i}$ are mutually distinguishable (by the proof above). From strong independence, there are no correlations between the environments conditioned on system state $\ket{i}$, hence the conditional state $\rho_{\mathcal{F}|i}$ is product.
We therefore have that $\rho_{\mathcal{SF}}=\sum_{i}p_{i}\ket{i}\bra{i}\otimes\rho_{\mathcal{E}_{1}|i}\otimes\cdots\otimes\rho_{\mathcal{E}_{F}|i}$, \emph{i.e.}, the extended spectrum broadcast structure.\hfill$\blacksquare$

\section{System objectivity without strong independence}

We argue that objectivity does not require strong independence\textemdash that the environments can be correlated without detriment to system state objectivity.

\begin{example}
Consider the following system-environment state:
\begin{align}
\rho_{\mathcal{S}\mathcal{E}_1\mathcal{E}_2\ldots\mathcal{E}_N} &= p_1 \ket{1}_\mathcal{S}\bra{1} \otimes \dfrac{1}{2} \left( \bigotimes_{k=1}^N\ket{1}_{\mathcal{E}_k}\bra{1}+\bigotimes_{k=1}^N\ket{2}_{\mathcal{E}_k}\bra{2}\  \right) + p_2 \ket{2}_\mathcal{S}\bra{2} \otimes \dfrac{1}{2} \left( \bigotimes_{k=1}^N\ket{3}_{\mathcal{E}_k}\bra{3}+\bigotimes_{k=1}^N\ket{4}_{\mathcal{E}_k}\bra{4}  \right),
\end{align}
which satisfies quantum Darwinism and has bipartite spectrum broadcast structure relative to reduced states $\rho_{\mathcal{S}\mathcal{E}_k}$, $k=1,\ldots,N$. The system state is objective: observers to environments $\mathcal{E}_k$ can independently determine the state of system without (Bohr-)disturbing it by using a POVM with elements $P_1 \propto \ket{1}\bra{1} + \ket{2}\bra{2}$ and $P_2 \propto  \ket{3}\bra{3}+\ket{4}\bra{4}$ for example.  However, strong independence is not satisfied. Hence, if an observer had a finer POVM of $\{\ket{i}\bra{i} \}_{i=1,2,3,4}$, then they could figure out extra information about other subenvironments. Objectivity of the system does not require objectivity of the environments in this example.
\end{example}

Furthermore, the subenvironments can also be entangled, such as the following:

\begin{example}
The following system-environment state satisfies system-objectivity (without subenvironment objectivity):
\begin{align}
\rho_{\mathcal{S}\mathcal{E}_1\mathcal{E}_2\ldots\mathcal{E}_N} &= p_1 \ket{1}_\mathcal{S}\bra{1} \otimes \dfrac{1}{2} \left( \ket{1\cdots1}_{\mathcal{E}_1\mathcal{E}_2\ldots\mathcal{E}_N} +\ket{2\cdots2}_{\mathcal{E}_1\mathcal{E}_2\ldots\mathcal{E}_N}  \right) \left( \bra{1\cdots1}+\bra{2\cdots2}  \right) \nonumber\\
& \phantom{=}+ p_2 \ket{2}_\mathcal{S}\bra{2} \otimes \dfrac{1}{2} \left( \ket{3\cdots3}_{\mathcal{E}_1\mathcal{E}_2\ldots\mathcal{E}_N} +\ket{4\cdots4}_{\mathcal{E}_1\mathcal{E}_2\ldots\mathcal{E}_N}  \right) \left( \bra{3\cdots3}+\bra{4\cdots4}  \right).
\end{align}
\end{example}

\section{Example of Horodecki \emph{et al} (2015) \cite{Horodecki2015}}\label{app:Horodecki}

\citet{Horodecki2015} provide an example where traditional quantum Darwinism emerges even when the system-environment state was clearly entangled. They give the following bipartite state:
\begin{align}
\rho_{\mathcal{SE}} & =pP_{\left(a\ket{00}+b\ket{11}\right)}+\left(1-p\right)P_{\left(a\ket{10}+b\ket{01}\right)},
\end{align}
where $P_{\ket{\psi}}=\ket{\psi}\bra{\psi}$, $p\neq1/2$, $a=\sqrt{p}$ and $b=\sqrt{1-p}$. The reduced system state is $\rho_{\mathcal{S}}=\Tilde{p}\ket{0}\bra{0}+\left(1-\Tilde{p}\right)\ket{1}\bra{1}$, where $\Tilde{p} = pa^2 + (1-p)b^2 = p^2 + (1-p)^2$ hence $H\left(\mathcal{S}\right)=H\left(\Tilde{p},1-\Tilde{p} \right)$. The reduced state of environment is $\rho_{\mathcal{E}}=p\ket{0}\bra{0}+\left(1-p\right)\ket{1}\bra{1}$, hence $H\left(\mathcal{E}\right)=H\left(p, 1-p\right)$, which is equal to the von Neumann entropy of the total state: $H\left(\mathcal{SE}\right)=H\left(p, 1-p\right)$. Hence the quantum mutual information is $I\left(\mathcal{S}:\mathcal{E}\right)=H\left(\mathcal{S}\right)$, \emph{i.e.}. quantum Darwinism is satisfied for all $p$.

However, with strong Quantum Darwinism, we must also look at the post-measurement state in the $\left\{ \ket{0},\ket{1}\right\} $ basis,
\begin{equation}
\begin{split}\rho_{\mathcal{S}^{\Pi}\mathcal{E}} & =\Tilde{p}\ket{0}\bra{0}\otimes\left(p^2\ket{0}\bra{0}+\left(1-p\right)^2\ket{1}\bra{1}\right)+\left(1-\Tilde{p}\right)\ket{1}\bra{1}\otimes\dfrac{1}{2}\left(\ket{0}\bra{0}+\ket{1}\bra{1}\right),
\end{split}
\end{equation}
which gives Holevo information
\begin{align}
\chi\left(\mathcal{S}^{\Pi}:\mathcal{E}\right) & =H\left(p\ket{0}\bra{0}+\left(1-p\right)\ket{1}\bra{1}\right)-\Tilde{p}H\left(\dfrac{1}{\Tilde{p}}\left( p^2\ket{0}\bra{0}+\left(1-p\right)^2\ket{1}\bra{1} \right)\right)-\left(1-\Tilde{p}\right)H\left(\dfrac{1}{2}\left(\ket{0}\bra{0}+\ket{1}\bra{1}\right) \right)\\
 & =H\left(p, 1-p\right)-\Tilde{p}H\left(\dfrac{p^2}{\Tilde{p}},\dfrac{(1-p)^2}{\Tilde{p}}\right) - (1-\Tilde{p})H(\dfrac{1}{2},\dfrac{1}{2}),
\end{align}
which is generally $\chi\left(\mathcal{S}^{\Pi}:\mathcal{E}\right)\neq H\left(\mathcal{S}\right)$ unless $p=0$ or $p=1$. Hence, strong Quantum Darwinism is not satisfied in general, and this is consistent with the correct conclusion that the system is not objective.

\end{document}